\documentclass[12pt]{article}
\usepackage{graphicx}
\usepackage[bookmarksnumbered,colorlinks,plainpages]{hyperref}

\begin{document}

\title{Fourier's law and maximum path information}

\author{Q.A. Wang\\
{\it Institut Sup\'erieur des Mat\'eriaux et M\'ecaniques Avanc\'es}, \\
{\it 44, Avenue F.A. Bartholdi, 72000 Le Mans, France}}

\date{}

\maketitle

\begin{abstract}
By using a path information defined for the measure of the uncertainty of instable
dynamics, a theoretical derivation of Fourier's law of heat conduction is given on
the basis of maximum information method associated with the principle of least
action.
\end{abstract}


\section{Introduction}
The Fourier's law of heat conduction, very well tested in experiments with solids,
liquids and gases, is given by
\begin{eqnarray}                                            \label{++1}
\textbf{J}(x)=-\kappa\nabla_x T(x)
\end{eqnarray}
where $\bf{J}(x)$ is the heat flux, $T(x)$ the temperature at a position $x$ (in
general a vector) of the ordinary space and $\kappa$ the thermal conductivity. This
law is valid not only for the case of steady process, but also for the case where
temperature varies in time. The theoretical derivation of this experimental law for
fluids and solids is a long history and thus far remains a challenge for theorists
of nonequilibrium thermostatistics\cite{Bonetto}. Most of the past efforts for
deriving this law were focalized on special models like harmonic or anharmonic
crystals at steady state and local equilibrium\cite{Bonetto2}. A general derivation
is still missing.

In this work, we will try to give a generic derivation of this law based on maximum
information principle. The information we address in this work is a quantification
of the uncertainty of dynamical process. It is given by Shannon
formula\cite{Shannon}
\begin{eqnarray}                                            \label{++2}
H=-\sum_{k=1}^vp_k\ln p_k
\end{eqnarray}
with respect to certain probability distribution $p_k$ of that process and the index
$k$ is summed over all the possible evolutions of the system of interest in its
phase space $\Gamma$. A phase volume $\Omega$ occupied by the system in $\Gamma$
space can be partitioned into $v$ cells of volume $s_i$ with $i=1,2,...v$ in such a
way that $s_i\cap s_j=\emptyset$ ($i\neq j$) and $\cup_{i=1}^vs_i=\Omega$. A state
of the system can be represented by a sufficiently small phase cell in coarse
graining way. The movement of a dynamical system is represented by its trajectories
(in the sense of classical mechanics) in $\Gamma$ space.

When we look at a nonequilibrium system leaving an initial cell $a$ in the
$\Gamma$-space for some destinations, we can say that, if the motion of the system
is regular, there will be only one possible trajectory from $a$ to a final cell $b$,
or in other words, there will be only a fine bundle of paths which track each other
between the initial and the final cells. These trajectories must minimize action
according to the principle of least action and have unitary probability. Any other
path should have zero probability. But if the dynamics is instable with strong
sensitivity to initial condition, the things are different. Two points
indistinguishable in the initial cell can separate from each other exponentially. So
from a given initial cell, there may be many possible final cells each having a
probability to be visited by the system. More than one paths between two cells are
possible (see Figure 1).

\begin{figure}[t] \label{f1}
\includegraphics[width=12cm,height=16cm]{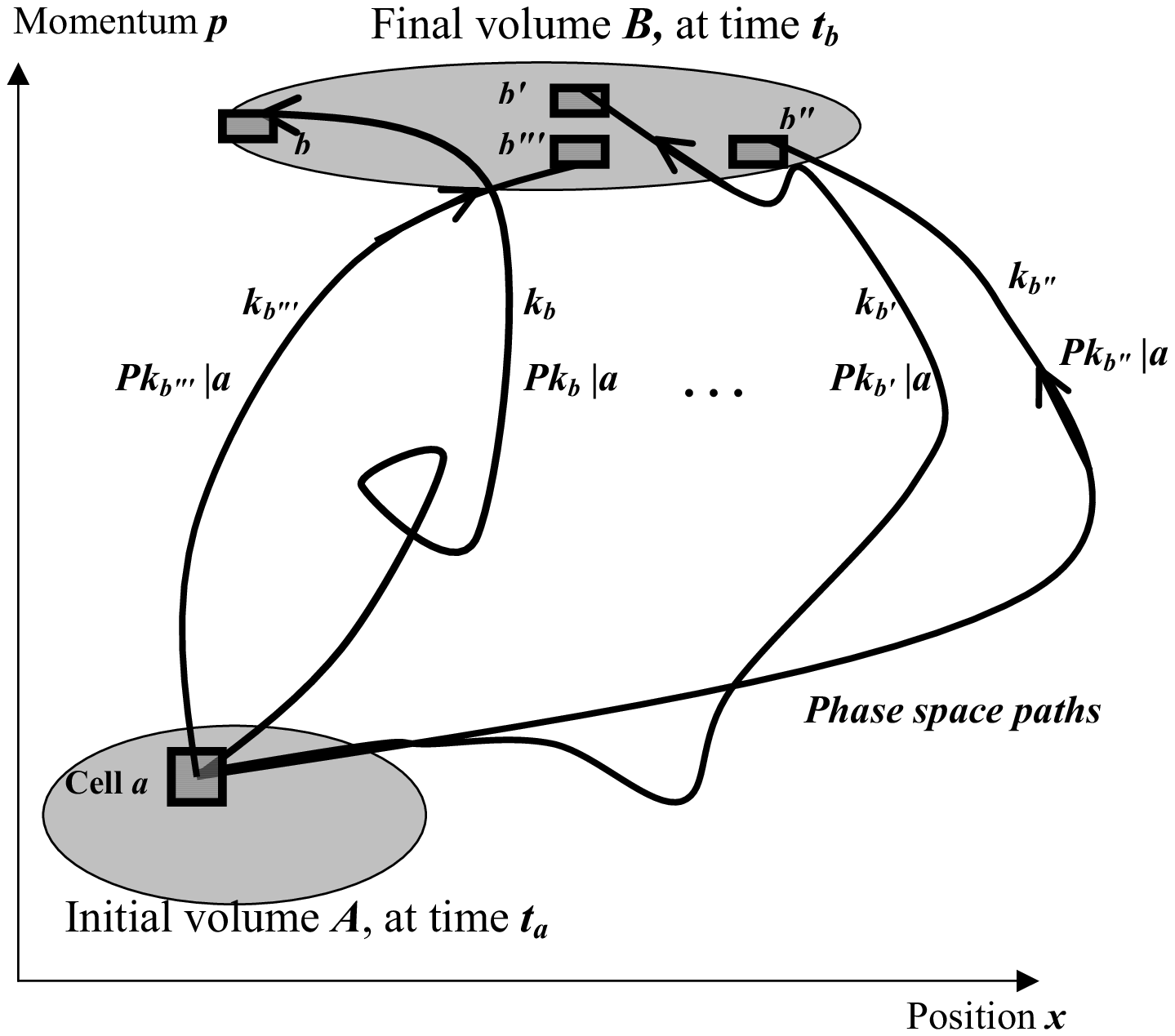}
\caption{The possible phase space paths for a system to go from a given cell $a$ to
different cells $b$ of a partition of the final phase volume $B$ during the time
$t_{ab}$, each having a probability $p_{k_b|a}$. More than one paths from $a$ to a
given cell $b$ are allowed so we may have $k_b=1,2 ... w_b$ where $w_b$ is the total
number of paths from $a$ to a cell $b$.}
\end{figure}

In this work, the above uncertainty is represented by a path information. This path
information will be proved to take its maximum when the most probable paths are just
the paths of least action. Then with the help of the model of Brownian motion, the
Fick's laws for diffusion and the Fourier law for heat conduction are derived.

\section{Maximum path information}
Let us consider an ensemble of large $N$ identical systems leaving the initial cell
$a$ for some destinations in the phase volume $B$ formed by the final phase points
occupied by the systems. The travelling time is $t_{ab}=t_b-t_a$. After $t_{ab}$,
all the phase points occupied by the systems are found in the volume $B$ partitioned
into cells labelled by $b$. We observe $N_{k_b}$ systems travelling along a path
$k_b$ leading to certain cell $b$. A path probability can be defined by
$p_{k_b|a}=N_{k_b}/N$ which is normalized by
\begin{eqnarray}                                            \label{c1xx}
\sum_{b}^{over B}\sum_{k_b=1}^{w_b}p_{k_b|a}=1.
\end{eqnarray}
where $w_b$ is the number of possible paths from $a$ to a given cell $b$ of the
volume $B$. We always suppose each path is characterized by its action $A_{k_b|a}$
defined for classical mechanical systems by
\begin{eqnarray}                                            \label{c5}
A_{k_b|a}=\int_{t_{ab}}L_{k_b}(t)dt
\end{eqnarray}
where $L_{k_b}(t)$ is the Lagrangian of the system at time $t$ along the path $k_b$.
The average action is given by
\begin{eqnarray}                                            \label{c1xxx}
A_a=\sum_{b}^{over B}\sum_{k_b=1}^{w_b}p_{k_b|a}A_{k_b|a}
\end{eqnarray}
The uncertainty concerning the choice of paths and final cells by the systems is
measured by the following Shannon information
\begin{eqnarray}                                            \label{c1x}
H_a=-\sum_{b}^{over B}\sum_{k_b}p_{k_b|a}\ln p_{k_b|a}
\end{eqnarray}
which we shall maximized under the constraints associated with Eq.(\ref{c1xx}) and
Eq.(\ref{c1xxx}) as follows

\begin{eqnarray}                                            \label{xc1x}
\delta (H_a+\alpha\sum_{b}^{over B}\sum_{k_b=1}^{w_b}p_{k_b|a}+\eta\sum_{b}^{over
B}\sum_{k_b=1}^{w_b}p_{k_b|a}A_{k_b|a})=0.
\end{eqnarray}
This leads to
\begin{eqnarray}                                            \label{c6x}
p_{k_b|a}=\frac{1}{Z_a}\exp[-\eta A_{k_b|a}],
\end{eqnarray}
where the partition function
\begin{eqnarray}                                            \label{cx6x}
Z_a=\sum_{b}^{over B}\sum_{k_b}\exp[-\eta A_{k_b|a}].
\end{eqnarray}
In path integral language, $Z_a$ can be given by\cite{Feynman}
\begin{eqnarray}                                            \label{cx6xx}
Z_a=\int_{over B}dp_bdx_b\int_a^b dpdx \exp\{-\eta
\int_{t_{ab}}[p\dot{x}-H(p,x)]dt\},
\end{eqnarray}
where $p=m\dot{x}$ is the momentum and $H(p,x)$ is the hamiltonian of the system.
Here the Lagrangian is given by $L_{k_b}(t)=p\dot{x}-H(p,x)$.

Putting Eq.(\ref{c6x}) back into Eq.(\ref{c1x}), we get
\begin{eqnarray}                                            \label{c7}
H_a=\ln Z_a+\eta A_a=\ln Z_a-\eta \frac{\partial}{\partial\eta}\ln Z_a.
\end{eqnarray}

It is proved that\cite{Wang04x} the distribution Eq.(\ref{c6x}) is stable with
respect to the fluctuation of action. It is also proved that if $\eta$ is positive,
Eq.(\ref{c6x}) is a least action distribution, i.e., the most probable paths are
just the paths of least action. On the contrary, if $\eta$ is negative, then
Eq.(\ref{c6x}) is a most action distribution which means that the most probable
paths maximize action. In any case, whatever the sign of the parameter $\eta$, the
most probable paths maximizing path information always correspond to extremum of
action ($\delta A_{k_b|a}=0$). {\it In other words, for instable dynamical process,
the method of maximum information must be used in order to derive correct
probability distributions just as the action principle must be used to derive the
correct trajectories for regular dynamics}.

\section{Transition probability of Brownian particles}
Now we consider a Markov diffusion process of an ensemble of identical particles of
mass $m$ idealized by Brownian motion. Suppose a certain path in Figure 1 along
which a Brownian particle of mass $m$ moves from $a$ to $b$ via a path $k$ which is
simplified by an intermediate cell $k_b$. Between the three cells $a$, $k_b$, and
$b$ situated in real space at $x_a$, $x_{k_b}$ and $x_b$, respectively, the particle
is free. The action $A_{k_b|a}$ of the particle from $a$ to $b$ can be calculated to
be
\begin{eqnarray}                                            \label{x9}
A_{k_b|a}=\frac{m(x_b-x_{k_b})^2}{2(t_b-t_{k_b})}
+\frac{m(x_{k_b}-x_a)^2}{2(t_{k_b}-t_a)},
\end{eqnarray}
Then from Eq.(\ref{c6x}), we have
\begin{eqnarray}                                            \label{c9}
p_{k_b|a} &=& \frac{1}{Z_{k_b}}
\exp\left[-m\eta\frac{(x_b-x_{k_b})^2}{2(t_b-t_{k_b})}\right]
\frac{1}{Z_{a}}\exp\left[-m\eta\frac{(x_{k_b}-x_a)^2}{2(t_{k_b}-t_a)}\right]
\end{eqnarray}
The separate normalization of the two factors of this distribution over all possible
positions $x_b$ and all possible paths between $a$ and $b$ gives
\begin{eqnarray}                                            \label{c9x}
Z_{k_b}Z_a= \left[\frac{2(t_b-t_k)}{m\eta}\right]^{d/2}
\left[\frac{2(t_k-t_a)}{m\eta}\right]^{d/2}
\end{eqnarray}
where $d$ is the dimension of the diffusion space.

On the other hand, according to the solution of diffusion equation\cite{Kubo}, the
transition probability for the particle to go from $a$ to $b$ via $k_b$ is
\begin{eqnarray}                                            \label{c8}
p_{k_b|a} &=& \frac{1}{[4\pi
D(t_b-t_{k_b})]^{d/2}}\exp\left[-\frac{(x_b-x_{k_b})^2}{4D(t_b-t_{k_b})}\right]\\\nonumber
&\times&\frac{1}{[4\pi
D(t_{k_b}-t_a)]^{d/2}}\exp\left[-\frac{(x_{k_b}-x_a)^2}{4D(t_{k_b}-t_a)}\right]
\end{eqnarray}
where $D$ is the diffusion coefficient. A comparison of Eq.(\ref{c8}) with
Eq.(\ref{c9}) leads to
\begin{eqnarray}                                            \label{xx9}
\eta=\frac{1}{2mD}
\end{eqnarray}
which implies $\eta$ is positive because $D>0$. So the highest information
distribution Eq.(\ref{c6x}) can be also called {\it least action distribution},
i.e., the most probable paths are just the paths of least action. The inverse of
this statement is: if the most probable paths minimize action, then the diffusion
constant $D$ must be positive.

The physical meaning of $\eta$ can be revealed if we use a general relationship
$D=\frac{l^2}{2d\tau}$\cite{Kubo} which leads to
\begin{eqnarray}                                            \label{xx9x}
\eta=\frac{d\tau}{ml^2},
\end{eqnarray}
where $l$ is the mean free path and $\tau$ the mean free time of the Brownian
particles.

\section{Fick's laws of diffusion}
With some mathematics, it can be proved that $p_{k_b|a}$ satisfies the following
dynamical equation:
\begin{eqnarray}                                            \label{c10}
\frac{\partial p_{k_b|a}}{\partial t} = D \Delta_x p_{k_b|a}
\end{eqnarray}
where $\Delta_x$ is the Laplacian of the diffusion space, $x=x_b$ and $t=t_b$. Let
$n_a$ and $n_b$ be the particle density at $a$ and $b$, respectively. The following
relationship holds generally
\begin{eqnarray}                                            \label{c10x}
n_b=\sum_{a}^{over A}\sum_{k_b}n_ap_{k_b|a}
\end{eqnarray}
which is valid for any $n_a$. This means
\begin{eqnarray}                                            \label{c10xx}
\frac{\partial n_b}{\partial t} = D \Delta_{x} n_b.
\end{eqnarray}
This is the second Fick's law of diffusion\cite{Kubo}. The first Fick's law
$\textbf{J}(x)=-D\nabla_x n_b(x)$ can be easily derived if we consider matter
conservation $-\frac{\partial n(x,t)}{\partial t}= \nabla_x\cdot \textbf{J}(x,t)$
where $\textbf{J}(x,t)$ is the flux of the particle flow.

The diffusion constant $D$ can be related to the partition function $Z_a$ by
combining Eq.(\ref{xx9}) and Eq.(\ref{c7}), i.e.,
\begin{eqnarray}                                            \label{cx10xx}
D=\frac{1}{2m}\frac{\partial A_a}{\partial H_a}=-\frac{1}{2m}\frac{\partial^2(\ln
Z_a)}{\partial H_a\partial\eta}.
\end{eqnarray}

\section{Fourier's law of heat conduction}
We consider a crystal idealized by a lattice of identical harmonic oscillators each
having an energy $e_k=N_\nu(x,t)h\nu$ where $h$ is the Planck constant, $\nu$ is the
frequency of a mode and $N_\nu(x,t)$ is the number of phonons of that mode situated
at $x$ at time $t$ in $x\rightarrow x+dx$ and $\nu\rightarrow \nu+d\nu$. Suppose
that there is no mass flow and other mode of energy transport in the crystal. Heat
is transported only through the phonon flow. The phonons of frequency $\nu$ diffuse
in the crystal lattice, among the lattice imperfections, impurities and other
phonons with in addition anharmonic effects\cite{Bonetto}, just like Brownian
particles of mass $m=h\nu/c^2$ having transition probability $p_{k_b|a}$. Let
$n_\nu=N_\nu/dx$ be the density of phonons which must satisfy
\begin{eqnarray}                                            \label{cx10x}
n_\nu(x_b,t)=\sum_{a}^{over A}\sum_{k_b}n_\nu(x_a,t)p_{k_b|a}
\end{eqnarray}
and also Eq.(\ref{c10xx}).

The total energy density $\rho(x,t)$ of phonons at $x$ and time $t$ is given by
\begin{eqnarray}                                            \label{xc10}
\rho(x,t)=\int_0^{\nu_m}h\nu n_\nu(x,t)dx\varrho(\nu)d\nu,
\end{eqnarray}
where $\varrho(\nu)d\nu dx$ is the mode number $x\rightarrow x+dx$ and
$\nu\rightarrow \nu+d\nu$, and $\nu_m$ is the maximal frequency of the lattice
vibration. This implies
\begin{eqnarray}                                            \label{xc10xx}
\frac{\partial \rho(x,t)}{\partial t} = D \Delta_{x} \rho(x,t)
\end{eqnarray}

A variation of energy density $\delta\rho(x,t)$ can be related to temperature change
$\delta T(x,t)$ by
\begin{eqnarray}                                            \label{11}
\delta\rho(x,t)=c\delta T(x,t)
\end{eqnarray}
where $c$ is the heat capacity per unit volume supposed constant everywhere in the
crystal. This leads to
\begin{eqnarray}                                            \label{c12}
\frac{\partial \rho}{\partial t} = \kappa\Delta_xT(x,t)
\end{eqnarray}
where $\kappa=Dc$ is the heat conductivity. This equation can be recast into
\begin{eqnarray}                                            \label{c12xx}
c\frac{\partial T(x,t)}{\partial t} = \kappa\Delta_xT(x,t)
\end{eqnarray}
which should be solved to give the evolution of temperature distribution due to the
heat flow. When a stationary state is reached, temperature is everywhere constant,
i.e., $\frac{\partial T(x,t)}{\partial t}=\Delta_xT(x,t)=0$. So the temperature
distribution is given by $\nabla_x T(x)=constant$.

Now considering the energy conservation in an elementary volume between $x$ and
$x+dx$ in which we have $-\frac{\partial \rho(x,t)}{\partial t}= \nabla_x\cdot
\textbf{J}(x,t)$, Fourier's law of heat conduction follows
\begin{eqnarray}                                            \label{c12x}
\textbf{J}(x)=-\kappa\nabla_x T(x)
\end{eqnarray}
where the time variable $t$ is removed since this law is independent of whether $J$
and $T$ vary in time.

\section{Concluding remarks}
A path information is defined in connection with different possible paths of
dynamical system moving in its phase space from the initial cell towards different
final cells. On the basis of the assumption that the paths are characterized by
their actions, we show that the maximum path information leads to an exponential
path probability distribution of action called least action distribution meaning
that the most probable paths are just the paths of least action. With the help of
the model of Brownian motion, we show that, from this least action distribution,
Fick's laws of diffusion and Fourier's law of heat conduction can be derived in a
general way without assumptions about the state of the transport.

\end{document}